\documentclass[aps,prd,twocolumn,superscriptaddress]{revtex4-2}

\usepackage{amssymb}
\usepackage{amsmath}
\usepackage{graphicx}
\usepackage{enumerate}
\usepackage{mathtools}
\usepackage{color}
\usepackage{bbold}
\usepackage{subfigure}
\usepackage{slashed}
\usepackage[dvipsnames,usenames]{xcolor}
\usepackage{bm}
\begin{document}


\title{Neutrino many-body correlations}


\author{Lucas Johns}
\email[]{ljohns@lanl.gov}
\affiliation{Departments of Astronomy and Physics, University of California, Berkeley, CA 94720, USA}
\affiliation{Theoretical Division, Los Alamos National Laboratory, Los Alamos, NM 87545, USA}

\begin{abstract}
This paper responds to suggestions that the standard approach to collective neutrino oscillations leaves out potentially important quantum many-body correlations. Arguments in favor of this idea have been based on calculations that, on close scrutiny, offer no evidence either way. Inadequacies of the usual quantum-kinetic formalism are not currently supported by the literature.
\end{abstract}

\maketitle

\section{Introduction \label{sec:intro}}

Neutrino propagation in matter is a subject with various applications across particle physics, astrophysics, and cosmology. It also has a long history of raising subtle questions pertaining to quantum mechanics \cite{kuo1989, raffelt1996, prakash2001, giunti2007, volpe2024}.

Foundational issues are at their sharpest in the most extreme environments, where neutrinos exist in such high densities that their self-interactions are nonnegligible. In this regime, the clean separation of system (neutrinos) and environment (thermal medium) is forfeited, and the out-of-equilibrium phenomenology expanded. Dogged investigation over decades has mapped out the various forms of collective neutrino oscillations that are enabled by self-interactions. Today the territories bear the labels of spectral swaps \cite{duan2006, duan2007, raffelt2007, fogli2007, dasgupta2009, friedland2010}, matter--neutrino resonances \cite{malkus2012, wu2016, vaananen2016, zhu2016}, slow instabilities \cite{kostelecky1993b, samuel1996, hannestad2006, raffelt2007b, mirizzi2013, raffelt2013, duan2015b}, fast instabilities \cite{sawyer2005, sawyer2009, sawyer2016, chakraborty2016c, izaguirre2017, capozzi2017, dasgupta2017, abbar2018, johns2020, bhattacharyya2020, nagakura2021c, richers2021b, morinaga2022, nagakura2022, padillagay2022}, and collisional instabilities \cite{johns2023, padillagay2022b, lin2023, xiong2023c}. 

Piecing together all of these phenomena is a prerequisite for making trustworthy predictions that involve neutrino transport in extreme astrophysical and cosmological settings. Researchers are currently racing to meet this open challenge.

Yet, at the same time, an idea is increasingly circulating that the main part of this research community is hurtling forward on rattling, uninspected wheels. At issue are the roles of quantum many-body correlations and the mean-field approximation. (Following convention, we will here use the descriptor \textit{many-body} to refer specifically to aspects that are neglected in the standard quantum-kinetic formulation of neutrino transport.) Questioning in this vein dates back many years \cite{pantaleone1992, pantaleone1992b, friedland2003, friedland2003b, bell2003, sawyer2004, friedland2006, balantekin2007, pehlivan2011}, but lately it has entered a new era of buzzing activity \cite{birol2018, cervia2019, roggero2021, roggero2021b, martin2022, xiong2022c, roggero2022b, lacroix2022, cervia2022, martin2023}. Reviews have recently taken stock of this line of inquiry \cite{patwardhan2023, balantekin2023c} and situated it in the context of fundamental-physics applications of quantum information theory and quantum simulation \cite{klco2022}.

Some of the very first dedicated studies, those carried out by Friedland and Lunardini, concluded that many-body effects are not a serious concern \cite{friedland2003, friedland2003b}. Curiously, the type of analysis used in those papers has been all but abandoned ever since. Subsequent work has almost exclusively adopted a different approach. The evolution under a second-quantized Hamiltonian is numerically solved for a fixed number of neutrinos, and the result is compared to the mean-field evolution. Deviations are discovered, which are then interpreted as evidence that mainstream calculations of neutrino flavor conversion are on unsure footing.

The objective here is to explain why such an interpretation is unjustified. A recent work by Shalgar and Tamborra expresses the same conclusion, coming from a somewhat different angle \cite{shalgar2023c}. For the most part we do not restate points that have already been made by those authors. A fuller picture of the relevant physics will emerge from considering the two papers together. We also, of course, encourage readers to consult the many-body literature itself. This is important to stress because we focus here only on the implications of many-body calculations for the accuracy of neutrino quantum kinetics. The papers we are responding to contain many ideas and innovations that do not strictly fall within this scope.

The critique below confronts the many-body literature at a basic level, taking issue with the choice of Hamiltonian (Sec.~\ref{sec:scattering}) and the comparison to mean-field results (Sec.~\ref{sec:meanfield}). It also attempts to clarify the significance of finite-$N$ effects ($N$ being the number of neutrinos) and to bring attention to the underappreciated insights of Refs.~\cite{friedland2003, friedland2003b} (Sec.~\ref{sec:number}). The paper concludes with some brief thoughts on the tricky issue of properly testing the quantum-kinetic theory of neutrino transport (Sec.~\ref{sec:forward}).

\section{Requantized coherent scattering \label{sec:scattering}}

Like most prior work on this topic, we assume for simplicity that neutrinos come in only two flavors. The starting point for many-body calculations is the Hamiltonian \cite{balantekin2007, pehlivan2011}
\begin{equation}
H_\textrm{MB} = \sum_{\mathbf{p}} \omega_\mathbf{p} \mathbf{B} \cdot \mathbf{J}_\mathbf{p} + \frac{\sqrt{2} G_F}{V} \sum_{\mathbf{p}, \mathbf{q}} \left( 1 - \mathbf{\hat{p}} \cdot \mathbf{\hat{q}} \right) \mathbf{J}_\mathbf{p} \cdot \mathbf{J}_\mathbf{q}. \label{eq:MBH}
\end{equation}
In this equation, $G_F$ is the Fermi constant, $V$ is the quantization volume, $\mathbf{B}$ is the mass basis vector, $\omega_\mathbf{p} = \delta m^2 / 2 | \mathbf{p} |$ is the vacuum oscillation frequency (written here in terms of the mass-squared splitting $\delta m^2$) of neutrinos with momentum $\mathbf{p}$, and $\mathbf{J}_\mathbf{p}$ is a flavor isospin vector. The sums are over all momenta.

Unlike $\mathbf{B}$, the isospin vectors are operators. They are constructed from the creation and annihilation operators $a_i^\dagger (\mathbf{p})$ and $a_i (\mathbf{p})$ associated with the neutrino field of mass $m_i$ ($i = 1,2$):
\begin{align}
&J^z_\mathbf{p} (t) = \frac{1}{2} \left( a_1^\dagger (t, \mathbf{p}) a_1 (t, \mathbf{p}) - a_2^\dagger (t, \mathbf{p}) a_2 (t, \mathbf{p}) \right), \notag \\
&J^x_\mathbf{p} (t) = \frac{1}{2} \left( a_1^\dagger (t, \mathbf{p}) a_2 (t, \mathbf{p}) + a_2^\dagger (t, \mathbf{p}) a_1 (t, \mathbf{p}) \right), \notag \\
&J^y_\mathbf{p} (t) = \frac{1}{2i} \left( a_1^\dagger (t, \mathbf{p}) a_2 (t, \mathbf{p}) - a_2^\dagger (t, \mathbf{p}) a_1 (t, \mathbf{p}) \right). \label{eq:isospin}
\end{align}
The operators in Eq.~\eqref{eq:isospin} generate the Lie algebra of SU(2), the symmetry group of two-flavor neutrino mixing. It is also sometimes convenient to work in terms of the raising and lowering operators
\begin{align}
&J^+_\mathbf{p} (t) = J^x_\mathbf{p} (t) + i J^y_\mathbf{p} (t) = a_1^\dagger (t, \mathbf{p}) a_2 (t, \mathbf{p}), \notag \\
&J^-_\mathbf{p} (t) =  J^x_\mathbf{p} (t) - i J^y_\mathbf{p} (t) = a_2^\dagger (t, \mathbf{p}) a_1 (t, \mathbf{p}).
\end{align}
As usual, the creation and annihilation operators satisfy the anticommutation relations
\begin{equation}
\lbrace a_i (t, \mathbf{p}), a^\dagger_j (t, \mathbf{q}) \rbrace = (2\pi)^3 \delta^3 (\mathbf{p} - \mathbf{q}) \delta_{ij}.
\end{equation}
We could (and in a moment will) include helicity labels as well, in which case the right-hand side would feature an additional Kronecker delta.

The pairing terms $( 1- \hat{\mathbf{p}} \cdot \hat{\mathbf{q}} ) \mathbf{J}_\mathbf{p} \cdot \mathbf{J}_\mathbf{q}$ that appear in $H_\textrm{MB}$ come from neutral-current four-neutrino weak interactions. We will give a sketch of this derivation momentarily, in service of showing what terms $H_\textrm{MB}$ is missing.

Many-body studies invariably compare their quantum results to those obtained using the mean-field approximation, which consists of the replacement
\begin{equation}
\mathbf{J}_\mathbf{p} \cdot \mathbf{J}_\mathbf{q} ~\longrightarrow~ \mathbf{J}_\mathbf{p} \cdot \langle \mathbf{J}_\mathbf{q} \rangle + \langle \mathbf{J}_\mathbf{p} \rangle \cdot \mathbf{J}_\mathbf{q} - \langle \mathbf{J}_\mathbf{p} \rangle \cdot \langle \mathbf{J}_\mathbf{q} \rangle. \label{eq:MFA}
\end{equation}
Angle brackets signify expectation values. Eq.~\eqref{eq:MFA} is the focus of Sec.~\ref{sec:meanfield}.

The many-body Hamiltonian in Eq.~\eqref{eq:MBH} has by now taken on a life of its own. Unfortunately, it is not representative of neutrino transport in our universe. It is a \textit{requantization} of the flavor-space part of the neutrino Hamiltonian derived from quantum kinetics.

The real ground level, of course, is the Standard Model Lagrangian plus neutrino mass. Most pertinently, the $Z$-mediated neutrino--neutrino interaction has Hamiltonian density
\begin{equation}
\mathcal{H}_{\textrm{int}}^{\nu\nu} = \frac{G_F}{\sqrt{2}} \sum_{\alpha, \beta} \left( \bar{\nu}_\alpha \gamma^\mu P_L \nu_\alpha \right) \left( \bar{\nu}_\beta  \gamma_\mu P_L \nu_\beta \right), \label{eq:Hintdensity}
\end{equation}
where $\alpha$ and $\beta$ are flavor labels and $P_L = (1 - \gamma_5)/2$ is the left-handed chiral projector. The neutrino fields and $\mathcal{H}_{\textrm{int}}^{\nu\nu}$ carry implicit $t$- and $\mathbf{x}$-dependence. The Hamiltonian of the full system is
\begin{equation}
H_{\textrm{int}}^{\nu\nu} (t) = \int d^3 \mathbf{x}~ \mathcal{H}_{\textrm{int}}^{\nu\nu} (t, \mathbf{x}). \label{eq:Hint}
\end{equation}
Expanded in helicity states and momentum plane waves, the neutrino field is
\begin{align}
\nu_i (t, \mathbf{x}) = \sum_{h} \int &\frac{d^3 \mathbf{p}}{(2\pi)^3}~ \bigg( a_{i,h} (t, \mathbf{p}) u_{i,h} (\mathbf{p}) \notag \\
+~&b_{i,h}^\dagger (t, -\mathbf{p}) v_{i,h} (t, -\mathbf{p}) \bigg) e^{i \mathbf{p} \cdot \mathbf{x}} \label{eq:fieldexp}
\end{align}
with helicity $h$ and Dirac spinors $u$, $v$. We have introduced creation and annihilation operators $b^\dagger$ and $b$ for antineutrinos and are now displaying the helicity labels that were suppressed in Eq.~\eqref{eq:isospin}.

After using the expansion Eq.~\eqref{eq:fieldexp} in Eq.~\eqref{eq:Hint} and evaluating the integral over $\mathbf{x}$, we find
\begin{align}
H_{\textrm{int}}^{\nu\nu} = &\frac{G_F}{\sqrt{2}} \sum_{i, j} \sum_{\lbrace h \rbrace} \int \frac{d^3 \mathbf{p}_1}{(2\pi)^3} \frac{d^3 \mathbf{p}_2}{(2\pi)^3} \frac{d^3 \mathbf{p}_3}{(2\pi)^3} \frac{d^3 \mathbf{p}_4}{(2\pi)^3} \notag \\
\times &(2\pi)^3 \delta^3 ( \mathbf{p}_1 + \mathbf{p}_2 - \mathbf{p}_3 - \mathbf{p}_4 ) \notag \\
\times &\bigg[ \big( a_{i,h_3}^\dagger (t, \mathbf{p}_3) a_{i,h_1} (t, \mathbf{p}_1) a_{j,h_4}^\dagger (t, \mathbf{p}_4) a_{j,h_2} (t, \mathbf{p}_2) \big) \notag \\
&~~\times\big( \bar{u}_{i,h_3} (\mathbf{p}_3) \gamma^\mu P_L u_{i,h_1} (\mathbf{p}_1) \big) \notag \\
&~~\times\big( \bar{u}_{j,h_4} (\mathbf{p}_4) \gamma_\mu P_L u_{j,h_2} (\mathbf{p}_2) \big) + \dots \bigg]. \label{eq:Hintexpanded}
\end{align}
The notation $\lbrace h \rbrace$ indicates sums over $h_1, \dots, h_4$. Additional terms are not written out explicitly because only one of them is needed to make our point. Readers seeking a more thorough dissection of $H_\textrm{int}^{\nu\nu}$ might consult Refs.~\cite{sigl1993, volpe2013, vlasenko2014, kartavtsev2015, froustey2020}

For a moment let us confine our attention to forward scattering with $\mathbf{p}_3 = \mathbf{p}_1 \equiv \mathbf{p}$. The delta function imposes $\mathbf{p}_4 = \mathbf{p}_2 \equiv \mathbf{q}$ and the products of creation and annihilation operators become factors like $\mathbf{J}_\mathbf{p} \cdot \mathbf{J}_\mathbf{q}$. For the spinor contractions we have, for example,
\begin{align}
\big( \bar{u}_{i,-}  (\mathbf{p}) \gamma^\mu P_L u_{i,-} (\mathbf{p}) \big) \big( \bar{u}_{j,-}  (\mathbf{q}) \gamma_\mu &P_L u_{j,-} (\mathbf{q}) \big) \notag \\
&= 1 - \hat{\mathbf{p}} \cdot \hat{\mathbf{q}},
\end{align}
where the $-$ subscript labels left-handed helicity. In this way we recover the usual Hamiltonian used in many-body analyses \cite{balantekin2007}.

The point we wish to make here is simply that $H_\textrm{int}^{\nu\nu}$ (Eq.~\eqref{eq:Hintexpanded}) contains other terms besides those associated with forward scattering. Those contributions to the Hamiltonian are arbitrarily excluded from $H_\textrm{MB}$ (Eq.~\eqref{eq:MBH}). In the complete Standard Model Hamiltonian, forward scattering has no special priority. Even interpreting $H_\textrm{int}^{\nu\nu}$ as consisting of distinct types of scattering processes is to assume something about the particular system being treated. The notion of scattering applies when interactions can be regarded as isolated events acting on asymptotic states. This picture is fundamental to the quantum-kinetic description of neutrino transport. In many-body calculations, however, field modes continuously interact with one another and become progressively more correlated. This behavior undermines a scattering-based view of the dynamics.

Why, then, does the Hamiltonian used in the literature on many-body collective oscillations single out $\mathbf{J}_\mathbf{p} \cdot \mathbf{J}_\mathbf{q}$ pairing, neglecting the vastly larger set of other terms? This could stem from confusion regarding the origin of forward scattering, which is not in any universal sense a stronger interaction than all the others. The coherent enhancement of forward scattering is, ironically, a product of a quantum-kinetic treatment, in which interactions are conceptually divisible into sequences of scattering events. Yet that \textit{is} the standard approach to collective neutrino oscillations, the very one whose accuracy is up for debate.

The many-body literature requantizes the lowest-order kinetic (\textit{i.e.}, classical) Hamiltonian. The author is aware of no physical rationale for doing so.

\section{The mean-field approximation \label{sec:meanfield}}

It has become a familiar refrain that conventional calculations of neutrino flavor evolution are based on the mean-field approximation in Eq.~\eqref{eq:MFA}. Common though it may be, that statement is incorrect, or at least gives the wrong impression. As we have already indicated, the standard approach is in fact quantum kinetics, which makes several assumptions in addition to the mean-field approximation---and that last approximation is not made by applying Eq.~\eqref{eq:MFA} to Eq.~\eqref{eq:MBH}.

To test the accuracy of quantum kinetics, one needs to study models in which its underlying assumptions are actually plausible. The many-body literature has overlooked this point and solved problems to which the conventional approach is not even expected to be well-suited.

Quantum kinetics is founded on scale separation. Let $E$ be the energy scale characteristic of neutrino propagation in some medium. In sites where collective oscillations take place, neutrinos are produced by thermal processes and therefore have energies comparable to the temperature of the medium. At this coarse level of analysis, the other relevant dimensionful quantities are neutrino mass $m$, Fermi coupling $G_F$, and spatiotemporal gradients $\partial_x$. For our purposes, we can take $m$ and the gradients of the medium to be vanishingly small. Then the most important parameter characterizing the problem is the dimensionless quantity $G_F E^2$. Letting $\Sigma$ be the neutrino self-energy scale, we have
\begin{equation}
\frac{\textrm{potential energy}}{\textrm{kinetic energy}} \sim \frac{\Sigma}{E} \sim G_F E^2 \ll 1. \label{eq:peke}
\end{equation}
In a supernova or neutron-star merger, where $E \sim 10$~MeV, this ratio is $\sim 10^{-12}$---very small indeed. The weakness of the interactions motivates the use of kinetic theory, in which the excitations are neutrino quasiparticles with small self-energy corrections.

Collisional processes incite transitions between quasiparticle states at second order in the coupling. The ratio of the interaction time $\tau_\textrm{int}$ to the mean time $\tau^\textrm{c}_\textrm{mfp}$ between collisions is
\begin{equation}
\frac{\tau_\textrm{int}}{\tau_\textrm{mfp}^\textrm{c}} \sim \frac{E^{-1}}{(G_F^2 E^5)^{-1}} = (G_F E^2 )^2 \ll 1.
\end{equation}
A comparison of $\tau_\textrm{int}$ to the mean time $\tau^\textrm{fs}_\textrm{mfp}$ between forward-scattering events would simply return us to Eq.~\eqref{eq:peke}. Whether considering forward scattering or collisions, any given event is far removed in time from the one that preceded it. There are short intervals during which correlations develop, but very long ones during which they decay. The hypothesis of molecular chaos is built on this logic.

However, the mean-field approximation in the form of Eq.~\eqref{eq:MFA} is too strong. It means that neutrinos cannot become correlated at all, even during their interaction. Quantum kinetics applies the mean-field approximation to perturbatively expanded equations of motion rather than to the full Hamiltonian of the system. In elaborating on this point, we will attempt to minimize unnecessary detail. We again cite Refs.~\cite{sigl1993, volpe2013, vlasenko2014, kartavtsev2015, froustey2020} for comprehensive derivations.

Define the one-body operator $F_{\alpha\beta} = a^\dagger_\beta a_\alpha$. (We are henceforth dropping momentum and helicity labels.) The usual density matrix $\rho$ is obtained by taking the expectation value: $\rho = \langle F \rangle$. The operator equation of motion is
\begin{equation}
i \frac{d F}{dt} = \left[ H, F \right], \label{eq:Feom}
\end{equation}
which has formal solution
\begin{equation}
F(t) = F(0) - i \int_0^t dt'~ \left[ H(t'), F(t') \right]. \label{eq:Fsoln}
\end{equation}
Repeatedly substituting Eq.~\eqref{eq:Fsoln} into Eq.~\eqref{eq:Feom} generates an expansion analogous to the Lippmann--Schwinger equation. After a single substitution,
\begin{equation}
i \frac{d F}{dt} = \left[ H(t), F (0) \right] - i \int_0^t dt'~ \left[ H(t), \left[ H(t'), F(t') \right] \right].
\end{equation}
To turn this into an equation of motion for $\rho$, we need to take the expectation value $\langle \dots \rangle$ of both sides. Since $H = H[F]$ depends on the neutrino field, the mean-field approximation is used to factorize terms like
\begin{equation}
\langle \left[ H[F], F \right] \rangle = \left[ \langle H[F] \rangle, \langle F \rangle \right] = \left[ H[\rho], \rho \right].
\end{equation}
But there is a second issue, which is the dependence on earlier times. Handling the system's memory of its past requires separate assumptions. In particular, the Markov approximation is used to extend the limits of the integral to infinity, erasing the memory. Then, in accord with the hypothesis of molecular chaos, the equation of motion is assumed to hold at all times, not just at $t \cong 0$. Crucially, the second-order term does not vanish. It gives rise to Boltzmann collision integrals.

The approximations used in deriving quantum kinetics are justified by the hierarchy of time scales
\begin{equation}
\tau_\textrm{int} \ll t \ll \tau_\textrm{mfp}^\textrm{fs} \ll \tau_\textrm{mfp}^\textrm{c}, \label{eq:hierarchy}
\end{equation}
where $t$ symbolizes the time elapsed during the evolution. The power of molecular chaos is that it extends the validity of the resulting equation of motion out to longer times than this hierarchy would seem to permit.

With respect to the first-order term, we have
\begin{equation}
\left[ H(t), \rho (0) \right] \cong \left[ H(t), \rho (t) \right]
\end{equation}
because of Eq.~\eqref{eq:hierarchy}. Thus, to first order, the quantum kinetic equation precisely matches the mean-field equation obtained from applying Eq.~\eqref{eq:MFA} to Eq.~\eqref{eq:MBH}. But kinetic theory does not advise using the first-order equation for time scales longer than $\tau_\textrm{mfp}^\textrm{fs}$. Beyond that scale, collisions become significant. We can regard these second-order terms as a model of how information loss leads to relaxation. In this sense, the $G_F^2$ terms are actually approximating the backreaction of many-body correlations on one-body expectation values.

In short, the comparisons that have been made between many-body and mean-field calculations do not weigh in on the validity of quantum kinetics. The application of quantum kinetics is based on the notion that the neutrino field in an astrophysical environment is well-described by an ensemble of particles (\textit{i.e.}, spatially localized excitations). This picture motivates the formal manipulations that are used to derive the quantum kinetic equation. In particular, the existence of an interaction time $\tau_\textrm{int}$---whose short duration is used to justify the Markov approximation---is usually thought to be related to the overlap time between any two particles in the medium. According to this logic, quantum kinetics is not necessarily expected to describe the evolution of a quantum state prepared in such a way that there are no localized excitations. These are the sorts of states whose evolution has been simulated thus far in many-body calculations. They have
\begin{equation}
\langle a_{i,h}^\dagger (t, \mathbf{p}) a_{i',h'} (t, \mathbf{p'}) \rangle = 0 \notag
\end{equation}
for all $\mathbf{p'} \neq \mathbf{p}$. As a result, the Wigner distribution function,
\begin{align}
&\rho_{ij} (t, \mathbf{x}, \mathbf{p}) = \notag \\
&~~~~~\int \frac{d^3 \mathbf{q}}{(2 \pi)^3} e^{i \mathbf{q} \cdot \mathbf{x}} \left\langle a_{i,h}^\dagger \left( t, \mathbf{p} - \frac{\mathbf{q}}{2} \right) a_{i',h'} \left( t, \mathbf{p} + \frac{\mathbf{q}}{2} \right) \right\rangle, \notag
\end{align}
has no dependence on $\mathbf{x}$. A description in terms of particles propagating freely in between brief interactions does not appear to be plausible for this highly restrictive class of states, and so quantum-kinetic evolution is doubtful from the beginning.

\section{Finite-number effects \label{sec:number}}

As noted in Sec.~\ref{sec:intro}, some of the earliest papers to take up the topic of many-body effects argued that they are not a concern. These works include the ones by Friedland and Lunardini \cite{friedland2003, friedland2003b}, which we emphasized previously, and a later one by Hannestad, Raffelt, Sigl, and Wong \cite{hannestad2006}, which briefly addresses quantum fluctuations. It is worth asking how the rest of the many-body literature has sought to explain the discrepancy between their findings and the conclusions drawn in the three papers just cited.

The usual account, as found in Refs.~\cite{bell2003, roggero2021, roggero2021b, patwardhan2023, balantekin2023c} and elsewhere, begins by observing that the calculations by Friedland and Lunardini neglected one-body terms and angle-dependent couplings in the Hamiltonian. These simplifying choices (the thinking goes) led the authors to find that the many-body terms in flavor-conversion probabilities are down by a factor of the neutrino number $N$ compared to the mean-field terms and are thus relatively insignificant in the $N \rightarrow \infty$ limit. More realistic models lead to more worrisome findings, with deviations between many-body and mean-field outcomes developing on time scales proportional to $\log N$ (\textit{e.g.}, \cite{roggero2021b}). Even if we take $N$ to have the enormous value of $10^{58}$, approximately the total number of neutrinos emitted by a core-collapse supernova, the logarithmic factor is not overwhelmingly large. It would appear that many-body effects---in particular those related to the $N \rightarrow \infty$ limit being an imperfect approximation---were erroneously deemed unimportant in Refs.~\cite{friedland2003, friedland2003b} because those studies were based on overly simplistic models. Ref.~\cite{hannestad2006}, for its part, drew on those analyses in arguing that quantum fluctuations do not qualitatively change the oscillation physics.

The account above does not mention that Refs.~\cite{friedland2003, friedland2003b} differ from the rest of the many-body literature in a more significant respect. The Friedland--Lunardini analysis is predicated on a perturbative expansion in the dimensionless parameter $g = \sqrt{2} G_F t / V$. (They use the symbol $a$.) It is appropriate here to take $t$ to be the interaction time and $V$ to be the volume of the interaction region. Approximating these using the de Broglie wavelength, we have $t \sim E^{-1}$ and $V \sim E^{-3}$, where $E$ is a typical neutrino energy, and therefore
\begin{equation}
g \sim \sqrt{2} G_F E^2 \ll 1. \label{eq:g}
\end{equation}
Importantly, this is not an expansion in the coupling simpliciter. If $t$ were a free parameter, it could always be set to a value large enough to make $g$ exceed unity, no matter how small the coupling. Like $\Sigma / E$ in Eq.~\eqref{eq:peke}, $g$ measures the ratio of the interaction energy to the kinetic energy. The estimates in Eqs.~\eqref{eq:peke} and \eqref{eq:g} show that realistic situations are deep inside the perturbative regime. By using perturbation theory in $g$, Refs.~\cite{friedland2003, friedland2003b} work in the regime where quantum kinetics is applicable and are thus able to run self-consistency tests on the theory.

To return to $\log N$ scaling, let us again consider the density matrix $\rho$ with components $\rho_{\alpha\beta} = \langle a^\dagger_\beta a_\alpha \rangle$. Suppose that the system consists of $N$ neutrinos all with flavor $e$. Assuming the system to be generic (\textit{i.e.}, not prepared in some special manner), we should more carefully say that
\begin{equation}
\varrho \equiv \rho_{ee} = N, ~ \rho_{xx} = 0 \label{eq:varrho0}
\end{equation}
only to lowest order in quantum fluctuations. Similarly, the off-diagonal elements are not strictly vanishing but rather should have
\begin{equation}
\delta \equiv \left| \rho_{ex} \right| = \left| \rho_{xe} \right| = c \sqrt{N}, \label{eq:delta0}
\end{equation}
where $c$ is an $\mathcal{O}(1)$ number.

Now suppose further that the system is unstable to the development of flavor coherence in the classical ($N \rightarrow \infty$) limit. Then, within the same limit, the linearized dynamics has a solution
\begin{equation}
\delta (t) = \delta (0) e^{\Omega t}, ~ \Omega > 0.
\end{equation}
Let $\tau$ be the time required for this instability to lift $\delta \varrho$ to a fixed fraction $f$ of the initial $\nu_e$ density:
\begin{equation}
\frac{\delta (\tau)}{\varrho (0)} = f. \label{eq:fraction}
\end{equation}
Then, using Eqs.~\eqref{eq:varrho0} through \eqref{eq:fraction}, we obtain
\begin{equation}
\tau = \Omega^{-1} \log \frac{\delta (\tau)}{\delta (0)} = \Omega^{-1} \log \frac{f\sqrt{N}}{c}
\end{equation}
and, for large $N$,
\begin{equation}
\tau \sim \frac{1}{2} \Omega^{-1} \log N. \label{eq:logN}
\end{equation}
From this simple calculation we are able to see how classical instabilities amplify quantum fluctuations on time scales proportional to $\log N$.

In the collisionless limit, classical neutrino flavor instabilities come in two kinds: slow and fast. It is now understood that both slow \cite{hannestad2006, duan2007b, raffelt2013b, johns2018} and fast \cite{johns2020, padillagay2022, fiorillo2023} collective oscillations are, in their simplest forms, equivalent to pendular motion in flavor space. In the mechanical picture, flavor instabilities correspond to a pendulum swinging down from an upright position. The perfectly inverted pendulum is stationary in the classical limit, so the significance of quantum fluctuations is that they push the pendulum ever so slightly away from the unstable fixed point.

Their importance is possibly diminished by the fact that neutrino mass mixing already serves this purpose. Let $\vartheta$ be the zenith angle of the pendulum, with $\vartheta = 0$ at perfect inversion. Considering the simple case where the pendulum is not spinning, the equation of motion is
\begin{equation}
\ddot{\vartheta} = \Omega^2 \sin \vartheta.
\end{equation}
For small $\vartheta$,
\begin{equation}
\vartheta (t) \cong \frac{\vartheta_0}{2} \left( e^{\Omega t} + e^{- \Omega t} \right), \label{eq:varthetasoln}
\end{equation}
where $\vartheta_0 = \vartheta (0)$. At sufficiently long times ($t \gtrsim \Omega^{-1}$), the second term in parentheses drops out of Eq.~\eqref{eq:varthetasoln}. Some particular angle $\vartheta_\tau$ is reached after a time
\begin{equation}
\tau \cong \Omega^{-1} \log \frac{2 \vartheta_\tau}{\vartheta_0}.
\end{equation}
Taking $\vartheta_0 \rightarrow 0$ then implies
\begin{equation}
\tau \sim - \Omega^{-1} \log \vartheta_0.
\end{equation}
The initial displacement $\vartheta_0$, which is a stand-in for mass mixing, represents classical fluctuations. The $\log N$ scaling of Eq.~\eqref{eq:logN} does not change anything qualitatively in the early-time evolution. The question of the relative importance of quantum fluctuations and vacuum mixing for seeding instability is a quantitative one.

\section{Discussion \label{sec:forward}}

We have made three main points in this paper.

First (Sec.~\ref{sec:scattering}), we observed that the usual Hamiltonian used in the many-body literature, Eq.~\eqref{eq:MBH}, is actually a requantization of the coherent forward-scattering Hamiltonian derived from quantum kinetics. It does not give a more fundamental description of the dynamics than quantum kinetics does. It comes from an artificial restriction of the Standard Model weak-interaction Hamiltonian.

Second (Sec.~\ref{sec:meanfield}), we noted that the standard approach to collective neutrino oscillations does not rely on the mean-field approximation in the way that the many-body literature implies. Studies have typically asked whether Eq.~\eqref{eq:MFA} is a good approximation when applied to Eq.~\eqref{eq:MBH}. The real question is whether quantum kinetics is an accurate framework for treating neutrino transport. Of all the papers assessing the use of Eq.~\eqref{eq:MFA}, not one provides evidence that could guide us in answering the latter question. The only papers that have challenged quantum kinetics have concluded that it is justified \cite{friedland2003, friedland2003b}. Furthermore, there are intuitive reasons to expect quantum kinetics to be reliable (\textit{e.g.}, the relevant scales are indeed very well separated). No reasons have been offered as to why kinetics would fail or need to be extended in the case of a neutrino gas, the weakly coupled system par excellence.

Third (Sec.~\ref{sec:number}), we saw how $\log N$ scaling appears due to classical instabilities. Although a factor of $\log N$ is not extremely large, its finiteness might only influence flavor-conversion outcomes in ways that are already accounted for by classical effects. In general, we should be cautious when searching for finite-$N$ effects by comparing to classical systems with restrictive symmetries. The analysis in Sec.~\ref{sec:number} highlighted the flavor-space symmetry that prevails when the mixing angle is taken to zero, but the point also applies to phase-space symmetries.

Testing the validity of neutrino quantum kinetics is a subtle matter. The theory is based on several physically motivated approximations. For a test to be meaningful, it needs to be performed on a model where the conditions motivating those approximations are actually realized. It is not self-evident that a microscopically homogeneous model ($\partial_{\mathbf{x}} \rho_{ij} = 0$) meets the criteria.

Ideally we would compare a quantum-kinetic solution to one obtained from a more complete description of the same system. But if we try to ascend to a more general theory, as most of the many-body literature has done, what it means to model the \textit{same} system becomes hazy. In a fully quantum treatment, should the quantization volume enclose the entire supernova? When in the history of the star do we begin the calculation? And what do we do about our nearly complete lack of microscopic information about the system?

An alternative strategy would be to compare neutrino quantum kinetics as currently formulated to a higher-order quantum-kinetic theory. This could involve checking, for example, whether outcomes change significantly when two-body correlators are included as dynamical variables. Unfortunately, even this test could be somewhat ambiguous. Expanding equations of motion in a hierarchy of $n$-body correlation functions is not the same as expanding a function in a convergent series. 

As challenging as it may be, ascertaining the reliability of the standard quantum-kinetic description of neutrino transport is an important goal for this field. The present study has located no evidence that should make us suspect unreliability---but it rules nothing out.

\begin{acknowledgments}
The author acknowledges valuable conversations with Sajad Abbar, Tehya Andersen, Vincenzo Cirigliano, Hiroki Nagakura, Georg Raffelt, and Alessandro Roggero. This work was supported by NASA Hubble Fellowship grant number HST-HF2-51461.001-A awarded by the Space Telescope Science Institute, which is operated by the Association of Universities for Research in Astronomy, Incorporated, under NASA contract NAS5-26555.
\end{acknowledgments}

\bibliography{all_papers}

\end{document}